\documentclass[twocolumn,showpacs,superscriptaddress, preprintnumbers , amsmath,amssymb,letterpaper]{revtex4}

\RequirePackage{lineno}

\usepackage{dcolumn}
\usepackage{bm}
\usepackage{amsmath}
\usepackage{multirow}
\usepackage{graphicx}
\usepackage{float}
\usepackage[colorlinks=true]{hyperref}

\begin{document}


 \newcommand{\re}{\mathop{\mathrm{Re}}}
 \newcommand{\im}{\mathop{\mathrm{Im}}}
 \newcommand{\D}{\mathop{\mathrm{d}}}
 \newcommand{\I}{\mathop{\mathrm{i}}}
 \newcommand{\E}{\mathop{\mathrm{e}}}
 \newcommand{\unite}[2]{\mbox{$#1\,{\rm #2}$}}
 \newcommand{\myvec}[1]{\mbox{$\overrightarrow{#1}$}}
 \newcommand{\mynor}[1]{\mbox{$\widehat{#1}$}}
 \newcommand{\rmsemit}{\mbox{$\tilde{\varepsilon}$}}
 \newcommand{\mean}[1]{\mbox{$\langle{#1}\rangle$}}


\preprint{FERMILAB-PUB-13-323-APC}
\date{\today}
\title{First Operation of  an Ungated Diamond Field-Emission Array Cathode \\
 in a L-Band Radiofrequency Electron Source}

\author{P. Piot} \affiliation{Northern Illinois Center for
Accelerator \& Detector Development and Department of Physics,
Northern Illinois University, DeKalb IL 60115,
USA} \affiliation{Accelerator Physics Center, Fermi National
Accelerator Laboratory, Batavia, IL 60510, USA}
\author{C. A. Brau}  \affiliation{Department of Physics and Astronomy, Vanderbilt University, Nashville, TN 37235, USA}
\author{B. K. Choi} \affiliation{Department of Electrical Engineering and Computer Science, Vanderbilt University, Nashville, TN 37235, USA} 
\affiliation{Vanderbilt Institute of Nanoscale Science and Engineering, Vanderbilt University, Nashville, TN 37235, USA} 
\author{B. Blomberg} \affiliation{Northern Illinois Center for
Accelerator \& Detector Development and Department of Physics,
Northern Illinois University, DeKalb IL 60115, USA}
\author{W. E. Gabella} \affiliation{Department of Physics and Astronomy, Vanderbilt University, Nashville, TN 37235, USA}
\author{B. Ivanov} \affiliation{Department of Physics and Astronomy, Vanderbilt University, Nashville, TN 37235, USA}
\author{J. Jarvis} \affiliation{Advanced Energy Systems Inc., Medford, NY 11763, USA}
\author{M. H. Mendenhall} \affiliation{Department of Physics and Astronomy, Vanderbilt University, Nashville, TN 37235, USA}
\author{D. Mihalcea} \affiliation{Northern Illinois Center for
Accelerator \& Detector Development and Department of Physics,
Northern Illinois University, DeKalb IL 60115,
USA}
\author{H. Panuganti} \affiliation{Northern Illinois Center for
Accelerator \& Detector Development and Department of Physics,
Northern Illinois University, DeKalb IL 60115,
USA}
\author{P. Prieto} \affiliation{Accelerator Division, Fermi National
Accelerator Laboratory, Batavia, IL 60510, USA}
\author{J. Reid} \affiliation{Accelerator Division, Fermi National
Accelerator Laboratory, Batavia, IL 60510, USA}

\begin{abstract}
We report on the first successful operation of a field-emitter-array cathode in a conventional L-band radio-frequency electron source. The cathode consisted of an array of $\sim 10^6$ diamond diamond tips on pyramids. Maximum current on the order of 15~mA were reached and the cathode did not show appreciable signs of fatigue after weeks of operation. The measured Fowler-Nordheim characteristics, transverse beam density, and current stability are discussed. Numerical simulations of the beam dynamics are also presented. 
\end{abstract}

\pacs{ 29.27.-a, 41.85.-p,  41.75.Fr}
\maketitle

%
%
%
Over the past years, field-emission (FE) electron sources have been the subject of intense investigations due to several advantages they offer over photoemission and thermionic sources. The main advantages of FE sources stem from their ability to produce very low-emittance bunched beams, their capability to generate high-average current beams, and the absence of requirement for an auxiliary  laser system. A single-tip FE cathode emits electrons from a very small transverse area and can therefore produce beams with extremely small, near quantum-degenerate, transverse emittances~\cite{jensen,brauFE}. When arranged as large arrays, field-emission-array (FEA) cathodes can provide substantial average currents~\cite{nature} to the detriment of emittance which then scales linearly with the FEA macroscopic radius~\cite{psi,jarvis2012}. 

Pulsed field-emission occurs when a FE cathode experiences a time-dependent field, e.g.,  when located in a resonant radiofrequency (RF) cavity. Taking the example of a cylindrical-symmetric resonant pillbox cavity operating on the TM$_{010}$ mode with axial electric field $E_z(r=0,z,t)=E_0\cos(2\pi f  t )$, where $f$ and $E_0$ are respectively the field frequency and peak amplitude, field-emitted bunches have a root-mean-square (rms) duration  $\sigma_t \simeq  \omega^{-1} [\beta_e E_0 /B(\phi)]^{1/2}$ where $\omega\equiv 2\pi f$. The latter pulse duration is obtained by taking the current density  to follow the Fowler-Nordheim's (F-N) law~\cite{fn}  $j(t)=A(\phi) \beta_e^2 E(t)^2 \exp[-B(\phi)/(\beta_e E(t))]$ where $A(\phi)$ and $B(\phi)$ are functions of the work function $\phi$ of the cathode material and $\beta_e$ is a field-enhancement factor~\cite{FN,FN2}. Nominally, the bunch rms duration is a significant fraction of the RF field period typically resulting in beams with large energy spread. This limitation can however be circumvented by exposing the FE cathode to superimposed electromagnetic fields operating at harmonic frequencies with properly tuned relative phases and amplitudes. A practical implementation of this technique consists in a RF gun supporting two harmonic modes with axial electric fields~\cite{lewellen}.  \\

In this letter we report on the first operation of a diamond FEA (DFEA) cathode in a conventional L-band RF gun nominally operated with a Cesium Telluride (Cs$_2$Te) photocathode. The DFEA is composed of ungated diamond pyramids which have proven to be rugged.  Depending on the size and pitch of the pyramids, tests under DC voltages have showed field emission to begin at macroscopic fields $E_0\simeq 5$~MV/m, and peak currents per tip as high as 15~$\mu$A has been obtained~\cite{fea}.  

\begin{figure}[hhhhh!!!!!!!!!!!!]
\centering
\includegraphics[width=0.48\textwidth]{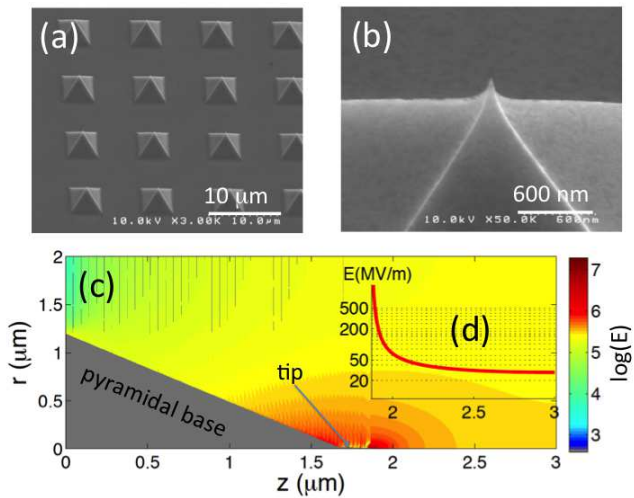}
\caption{Electron-microscope photographs of the DFEA pattern (a) and close up of one tip (b). Simulated electrostatic field (c) for a $E_0=30$~MV/m applied macrosopic  field. Inset (d) shows the on-axis ($r=0$) E-field amplitude from the FE tip ($z=1.8$~$\mu$m) to $z=3$~$\mu$m. }\label{fig:fea}
\end{figure}

The geometry of the DFEA cathode used for the experiment reported below appears in Fig.~\ref{fig:fea}(a,b). It consists of an array of $\sim 1000\times 1000$ diamond tips on pyramids with their extremities separated by $\sim 10$~$\mu$m. The typical pyramid base is  $\sim 4$~$\mu$m and the radius of curvature of the tip is on the order of 10~nm. Electrostatic simulations performed with the finite-difference element program {\sc poisson}~\cite{poisson}, indicate that local fields in excess of $\sim 0.9$~GV/m are achieved at the tip when subjected to a macroscopic field $E_0=30$~MV/m (corresponding to $\beta_e\sim 30$); see Fig.~\ref{fig:fea}(c,d). 

The cathode pattern was formed using a mold-transfer process whereby chemical vapor deposited (CVD) diamond is grown in sharpened silicon molds~\cite{dfea}. Using various techniques, DFEAs can be produced with single, double, or quadtip emitters. A variety of growth recipes are used to achieve a desired combination of $sp^2$ and $sp^3$ carbon,  and nitrogen content. The DFEAs are brazed on a molybdenum substrate.  \\

\begin{figure}[hhhhh!!!!!!!!!!!!]
\centering
\includegraphics[width=0.46\textwidth]{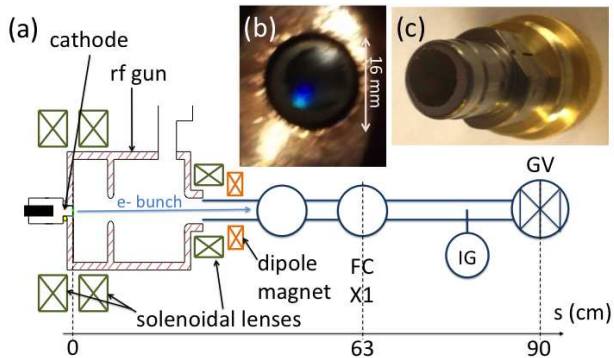}
\caption{Experimental setup used to perform field-emission studies (a) . The legend is as follows: "FC": Faraday cup, "IG" ion gauge, "X1": scintillating screen for transverse beam density measurement.The insets are photograph of the DFEA cathode in its final position in the RF gun (b), and of the cathode holder with diamond coating on its font surface before insertion in the RF gun (c).}\label{fig:setup}
\end{figure}

The DFEA cathode was  located on the back plate of a  1.625-cell RF gun operating on the TM$_{010}$ $\pi$ mode at $f=1.3$-GHz. The RF gun is powered by a $\le 3$-MW peak-power klystron pulsed at 1~Hz~\cite{carneiro}. For the measurement presented below the RF macropulse width was set to 35~$\mu$s. The RF gun is surrounded by three solenoidal lenses that control the beam's transverse size. The charge and transverse distribution of the emitted electron beam can be measured with respectively a Faraday cup (FC) or scintillating screen (X1); see Fig.~\ref{fig:setup}. Both diagnostics are located at $z=0.63$~m from the cathode and are remotely insertable. A small corrector dipole located upstream of X1 can be used to measure the beam momentum. The X1 screen is imaged on a charged-coupled-device (CCD) optical camera. A variable iris and neutral-density filters are used to attenuate the emitted optical radiation and mitigate saturation of the CCD. The optical resolution of the imaging system is $\sim 100$~$\mu$m. A ion gauge (IG) located 0.5~m from the gun monitors the vacuum level in the section composed of the gun and diagnostics. Typical vacuum pressure levels in this beamline section are $\sim 1\times 10^{-9}$~Torr. The forward RF power, $P$, injected in the RF gun cavity is measured using a calibrated RF diode detecting the low power from a -60-dB directional coupler installed on the RF waveguide. The measured forward power can  be used to infer the peak electric field at the cathode surface via $E_0 \mbox{[MV/m]} \simeq 2.234\times 10^{2} \sqrt{P\mbox{[MW]}}$ where the RF-gun quality factor is taken to be $Q\simeq 2.3\times 10^4$ and the cathode is assumed to be flush with the RF-gun back plate.

The DFEA's substrate was brazed on a molybdenium cathode holder compatible with the load-lock insertion mechanism used in the RF gun; see Fig.~\ref{fig:setup}(c). The cathode holder was inserted in the RF gun and its position was adjusted to insure the gun resonant frequency remains at 1.3 GHz (as monitored with a spectrum analyzer). To insure electrical contact between the cathode holder and RF-gun-cavity walls, a Cu-Be spring is used. This spring is also found to be a source of spurious negligible field-emission current.

\begin{figure}[hhhhh!!!!!!!!!!!!]
\centering
\includegraphics[width=0.46\textwidth]{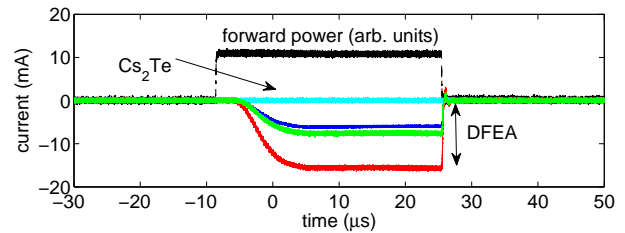}
\caption{Current traces recorded with the DFEA cathode for a 1.2 (blue), 1.3 (green) and 1.5~MW (red) forward power. The turquoise trace indicate the current observed when the nominal Cs$_2$Te photocathode is inserted in the RF gun. }\label{fig:current}
\end{figure}

A first series of measurement consisted in characterizing the "background'' field emission from the Cu-Be spring. For these studies a standard Cs$_2$Te photocathode -- a Mo cathode holder coated with Cs and Te layers -- was first used to quantify the spurious field-emission current as recorded by the FC. The DFEA cathode was then inserted and its current was measured for several forward power levels. Figure~\ref{fig:current}  compares the measured current and confirms that the spurious current induced by the Cu-Be RF spring did not appreciably contribute to our measurements as no significant emission current was detected  ($I \le 50$~$\mu$A within the noise level) with the Cs$_2$Te cathode. Maximum currents in excess of 15 mA were measured with the DFEA cathode. The current  waveform slow rising time for $t\in[-10,5]$~$\mu$s is consistent with the $e$-fold filling time of the RF gun $\tau = Q/\omega \simeq 3$~$\mu$s~\cite{carneiro}.  For the latter measurement the solenoidal lenses were turned off. 

\begin{figure}[hhhhh!!!!!!!!!!!!]
\includegraphics[width=0.46\textwidth]{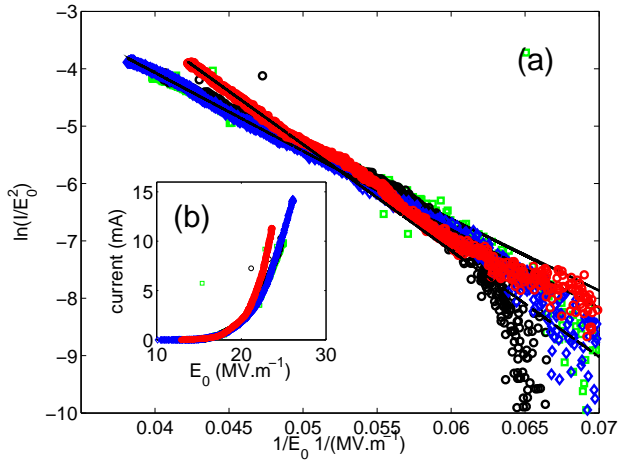}
\caption{F-N plots (a) for the nominal (green, blue, black traces) and retracted (red) cathode position. For the nominal cases, the black and blue traces were respectively taken 10 and 59 days later than the green trace). The inset (b) displays the current evolution versus macroscopic field amplitude. }\label{fig:FNplots}
\end{figure}

A sample of F-N characteristics recorded over several days of operations appear in Fig.~\ref{fig:FNplots}.  The curves all have  similar slopes at high fields [$\nu\equiv B(\phi)/\beta_e=129.5\pm 6.2$] for the cathode in its nominal position (with its emitting surface flush with the RF gun backplate). A somewhat different slope is observed ($\nu'= 213.2\pm 0.5$) when the cathode is retracted by $\sim 2$~mm. For all cases, the start-up macroscopic field is estimated to $E_0\simeq 20$~MV/m.  

\begin{figure}[hhhhh!!!!!!!!!!!!]
\centering
\includegraphics[width=0.46\textwidth]{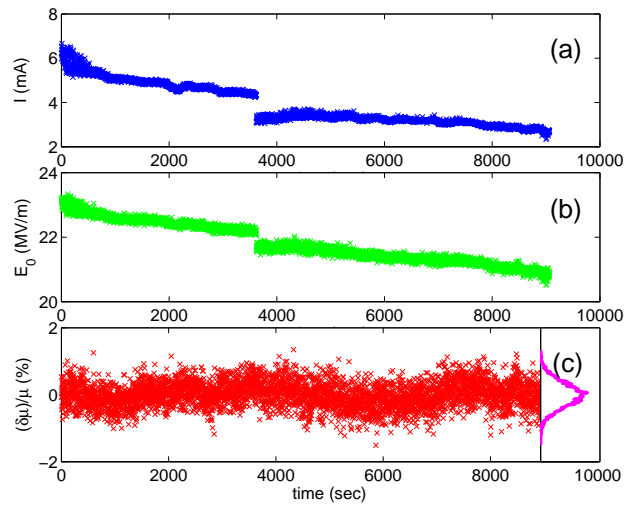}
\caption{Long-term stability studies for two cases of electric field value (initial setup points $E_0\simeq 23$~MV/m for $0\le t < 3800$~s and  $E_0\simeq 21.7$~MV/m  for $3800\le t < 9000$~s).  Evolution of the beam current (a), macroscopic field $E_0$ (b) and relative change in the F-N slope $\mu$ (c). A time-integrated histogram of $(\delta \mu)/{\mu}$ is shown in (c).}\label{fig:stab1}
\end{figure}

During the running period, the vacuum level was monitored and did not appreciably deteriorate ($\le 1.2 \times 10^{-9}$~Torr). The current was also recorded for long period of time (typically up to one to two hours); see Fig.~\ref{fig:stab1}. An apparent current drift was observed and tracked back to the  klystron power drifting in time. Accounting for the E-field drift by computing an ``instantaneous" F-N slope as ${ \mu}\equiv E_0\times \ln(I/E_0^2)$ indicates that the emission is stable with typical relative rms variation $\mean{\left[(\delta\mu)/\mu\right]^2}^{1/2} \simeq  0.37$\%. In practice correlated changes due to klystron-power drifts can easily be mitigated using a slow feedback loop.

\begin{figure}[hhhhh!!!!!!!!!!!!]
\centering
\includegraphics[width=0.46\textwidth]{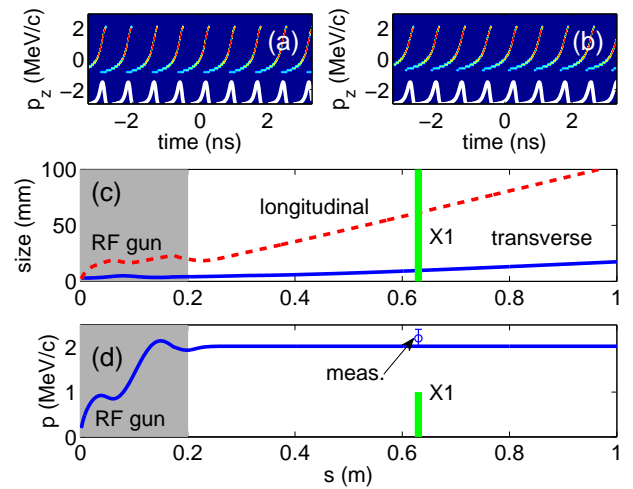}
\caption{Longitudinal phase spaces at $z=0.63$ (a) and $z=1.0$~m (b) from the DFEA cathode and beam's rms sizes (c) and mean momentum (d) evolutions along the beamline shown in Fig.~\ref{fig:setup}. The parameters are statistically evaluated over a single bunch (and not the full train). The data point labeled ``meas." in plot (d) corresponds to the measured beam momentum. \label{fig:astra}}
\end{figure}

To complement our measurement and gain further understanding of the beam dynamics of the field-emitted bunch we carry numerical simulation using the {\sc astra} program~\cite{astra}. In our simulations only the macroscopic features of the DFEA were incorporated: we model the emission as a source of electron with initial Gaussian temporal distribution with rms duration $\sigma_t=\frac{1}{2\pi f}\sqrt{E_0/\nu}$ (using the experimental values of $\nu$). The corresponding transverse shape is assumed to follow a radially-uniform distribution with radius $r=6$~mm. For the measured current, the charge per bunch is only a few pC so that space charge effects do not play a significant role.

Snapshots of the simulated longitudinal phase spaces $(t, p_z)$ associated to a bunch train (of $\sim 10$ bunches) at two axial location appears in Fig.~\ref{fig:astra}(a,b). The evolution of  the beam sizes and mean momentum are display in respectively Fig.~\ref{fig:astra}(c) and (d).  
\begin{figure}[hhhhh!!!!!!!!!!!!]
\centering
\includegraphics[width=0.46\textwidth]{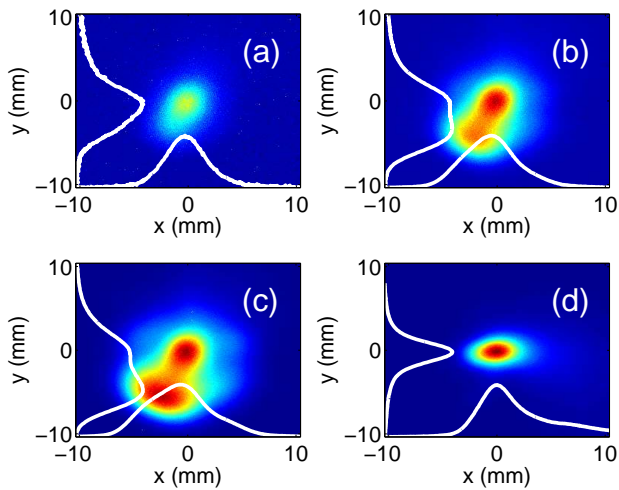}
\caption{Beam transverse density measured at X1 for three cases of macroscopic power $ E_0 \simeq 23.5$ (a), 25.5 (b) and 31~MV/m (c) with solenoids off and for a focused beam and $ E_0 =31$~MV/m (d). }\label{fig:spot}
\end{figure}

The low mean momentum [measured to be $\mean{p}=2.2\pm 0.1$~MeV/c consistent with simulations; see Fig.~\ref{fig:astra}(a,b)] and large momentum spread result in significant bunch lengthening due to non-relativistic effects occurring in drift spaces [bunch lengthening is $\propto  D/\gamma^2 \sigma_{\delta}$ where $D$ is the drift length, $\sigma_\delta$ the rms fractional momentum spread, $\gamma$ the Lorentz factor and $\beta\equiv \sqrt{1-1/\gamma^2}$]. Further acceleration or dispersive collimation could mitigate the bunch lengthening.  


Finally, the beam densities were measured at X1 for a sample of operating conditions; see Fig.~\ref{fig:spot}. When the solenoidal lenses are turned off, the beam size increases with the macroscopic field amplitude. This dependence is expected due to the radial force $F_r(r)\propto E_r(r) \simeq -(r/2)\partial E_z(z)/(\partial z)$ (in the paraxial approximation) experienced by an off-axis particle.  Despite (1) the large energy spread, and (2) the observed transverse distortion, the beam could nevertheless be refocused to a $\sim 1$-mm (rms) spot size; see Fig.~\ref{fig:spot}(d). Finally, we note that although unwanted in the present case, the impression of controlled geometric distortions on the cathode surface could in principle enable the generation of, e.g., bunch with shaped transverse distribution. \\

In summary we successfully demonstrated the operation of a DFEA cathode in a conventional L-band RF gun nominally designed to operate using photocathodes. These results represent a significant step toward the realization of robust laser-free compact light sources for industrial, medical, and defense applications. Further tests will focus on characterizing the emittance of the beams generated from smaller-area DFEA cathodes along with producing short bunch using gated cathodes currently under development~\cite{gated}.  \\

We are indebted to M. Church, E. Harms , E. Lopez,  J. Santucci, and V. Shiltsev for their support.  This work is supported by the DARPA Axis  contract AXIS N66001-11-1-4196 with Vanderbilt University, DOE Contracts No. DE-AC02-07CH11359 with the Fermi Research Alliance, LLC. and No. DE-FG02-08ER41532  with the Northern Illinois University.

\end{document}